\documentclass[12pt]{article}

\makeatletter
\@addtoreset{equation}{section}
\renewcommand*{\@makefnmark}{}
\makeatother
\newcommand*{\foot}[1]{\hbox{\mathsurround=0pt%
$^{\thefootnote}$}\addtocounter{footnote}{-1}%
\footnote{\hbox{\mathsurround=0pt $^{\thefootnote}$}#1}}

\usepackage{theorem}
\theorembodyfont{\upshape}

\newtheorem{theorem}{Theorem}[section]
\newtheorem{lemma}[theorem]{Lemma}

\newcounter{abc}[equation]

\newcommand*{\picdirrectory}{}
\newcommand*{\fig}[3]{
        \begin{figure}[!h!t]
        \begin{center}
        \includegraphics{\picdirrectory#1}
        \end{center}
        \caption{\footnotesize #2}
        \label{#3}
        \end{figure}}

\usepackage{amssymb,amsfonts}
\usepackage{graphics}

\makeatletter
\@addtoreset{equation}{section}
\makeatother

\def\ZZ{{\mathbb Z}}
\def\NN{{\mathbb N}}
\def\RR{{\mathbb R}}

\def\indic{{\bf 1}}

\newenvironment{eqs}{\begin{equation}\left\{\begin{array}{ll}}{\end{array}\right
.\end{equ
ation}
}
\newenvironment{eqe}{\begin{equation}\begin{array}{rl}}{\end{array}\end{equation
}}
\newenvironment{eqc}{\begin{equation}\begin{array}{c}}{\end{array}\end{equation}
}
\newcommand*{\be}{\begin{eqs}}
\newcommand*{\ee}{\end{eqs}}
\newcommand*{\bee}{\begin{eqe}}
\newcommand*{\eee}{\end{eqe}}
\newcommand*{\boe}{\begin{eqc}}
\newcommand*{\eoe}{\end{eqc}}
\newenvironment{proof}[1]{{\noindent{\sc 
Proof#1:}}}{\begin{flushright}$\square$\end{flushright}}

\newcommand*{\sumu}{\displaystyle\sum}

\renewcommand*{\epsilon}{\varepsilon}

\newcommand{\xit}{{\hat{\xi}}}
\newcommand{\etat}{{\hat{\eta}}}
\newcommand{\tot}{{\rightsquigarrow}}

\renewcommand{\picdirrectory}{bsnt.pic/}

\begin{document}
\pagestyle{myheadings} \markright{The Bak-Sneppen model}
\setlength{\textheight}{21cm}
\title{{\bf On the limit behaviour of the Bak-Sneppen evolution model}
\foot{\rm
{\it MSC 2000 subject classifications.} 60K35, 82B26, 92D15.}
\foot{\rm
{\it Key words and phrases.} Bak-Sneppen, species, fitness, evolution, 
interacting particle system,
           self-organised criticality, coupling, stationary distribution, phase 
transition, graph.}}
\author{Ronald Meester and Dmitri Znamenski\\
{\it Vrije Universiteit, Amsterdam}}
\maketitle

\begin{abstract}
One of the key problems related to the Bak-Sneppen evolution model on the circle
is to compute the limit distribution of the fitness at a fixed observation
vertex in the stationary regime, as the size of the system tends to infinity.
Simulations in~\cite{J} and~\cite{B} suggest that this limit distribution
is uniform on $(f,1)$, for some $f\sim2/3$. 
In this paper we prove that the mean of the fitness
in the stationary regime is bounded away from $1$, uniformly in the size 
of the system, thereby establishing the non-triviality of the limit behaviour.

The Bak-Sneppen dynamics can easily be defined on any finite connected graph.
We also present a generalisation of the phase-transition result in the context 
of an 
increasing 
sequence of such graphs. This generalisation covers the multi-dimentional 
Bak-Sneppen 
model
as well as the Bak-Sneppen model on a tree.

Our proofs are based on a `self-similar' graphical representation 
of the avalanches.
\end{abstract}

\section{Introduction}
The Bak-Sneppen model, introduced in \cite{BS}, has received a lot of
attention in the literature, see for instance \cite{B}, \cite{J} and \cite{BSE}.
In \cite{B}, it is described how Bak and Sneppen were looking for a simple
model which was supposed to exhibit evolutionary behaviour,
and which was also supposed to fall into the class of processes showing
self-organised critical behaviour. For physicists, self-organised critical
behaviour refers to power law decay of temporal and spatial quantities.
After a number of attempts, Bak and Sneppen arrived at the following simple 
process.

Consider a system with $N$ species. These species are represented by $N$
vertices on a circle, evenly spaced, say. Now each of these species is assigned
a so called {\em fitness}, a number between
0 and 1. The higher the fitness, the better chance of surviving the species
has. The dynamics of evolution is modelled as follows. Every discrete time
step, we choose the vertex with minimal fitness, and we think of the
corresponding species as disappearing completely. This species is then
replaced by a new one, with a fresh and independent fitness, uniformly
distributed on $[0,1]$. So far, the dynamics does not have any
interaction between the species, and does not result in an interesting process. 
Indeed, if we only replace the species with the lowest fitness, then it is easy 
to see that the system converges to a situation with all fitnesses equal to 1.
Interaction is
introduced by also replacing the two neighbours of the vertex with lowest
fitness by new species with independent fitnesses.
This interaction represents co-evolution of related species.
This neighbour interaction makes the model also very interesting
from a mathematical point of view.

It is simple to run this model on a computer. Simulations then
suggest the following behaviour, for large $N$ (see \cite{J} and \cite{B}
for simulation results). It appears that the one-dimensional marginals
are uniform (in the limit for $N \to \infty$) on $(f,1)$ for some $f$ whose
numerical value is supposed to be close to 2/3. This threshold value $f$
is the basis for self-organised critical behaviour, according to \cite{BS},
\cite{B} and \cite{J}, as follows. Since in the limit there is no mass below 
$f$, one
can look at so called {\em avalanches} of fitnesses below this threshold: 
starting
the counting at the moment all fitnesses are above $f$ and finishing the 
counting
at the first next moment all fitnesses are above $f$ again. The random
number of updates, for instance, counted this way, is supposed to follow a power 
law.
For this to make sense, the conjecture had better be true. This we have not been 
able
to prove. However, we have been able to prove the weaker result that the mean 
average 
fitness
in the stationary regime is bounded away from $1$, uniformly in the number
of vertices. Hence, in the limit, there is probability mass of the fitnesses 
below 1. 
Since it 
is not 
difficult to show that in the limit there can be no probability mass of the 
fitnesses 
below 1/3, our 
results establish that the one-dimensional marginals do not become trivial as $N 
\to \infty$. For a similar result in a 
discrete version of this process, see \cite{MZ}.

Let $F_N$  be the distribution function of the one-dimensional marginal in the 
stationary
regime, in the system with $N$ vertices. We will prove the following result.

\begin{theorem}
\label{P1}
If $q<1$ is close enough to $1$, then there exists $c_q>0$, independent of $N$,
such that
\boe
F_N(q)>c_q.
\eoe
\end{theorem}

In Section~\ref{G} and 5, we define the Bak-Sneppen dynamics on arbitrary finite 
connected graph,
and present a generalisation of our results for an increasing sequences 
of such 
graphs.

We remark that corresponding results for a mean-field version of the model are
quite simple to obtain, see e.g. \cite{deB} and \cite{J}. In the mean-field 
case, it is possible to prove, using essentially only combinatorics, that the 
one-dimensional marginals do indeed converge to a uniform distribution on
$(c,1)$, for some constant $c$ which depends on the characteristics of the 
model. 

In the next section we prepare for the proof of Theorem~\ref{P1}, by introducing 
the notion of an 
avalanche, and establishing
some monotonicity properties of the avalanches. The proof itself can
be found in Section 3.

\section{The self-similar graphical representation}
Let $\Lambda(N)=\{-N+1,\ldots,-1,0\}$ index the set of $N$ vertices on the 
circle,
so that $0$ and $-N+1$ are neighbours. We use negative indices to simplify 
notation
in the future.
We say that in the time interval $[n,n+d]$, an
{\em avalanche from threshold} $q\in[0,1]$ (also referred to as a $q$-avalanche)
with {\em origin} at $x\in\Lambda(N)$ and {\em duration} $d\ge1$ occurs if at 
time $n$,
$x$ is the vertex with minimal fitness, above threshold $q$,
and $n+d$ is the first moment after $n$ with all fitnesses again above $q$.
The {\em range set} of the $q$-avalanche is the collection of vertices
updated during the avalanche, and the {\em range} of the $q$-avalanche is the
number of different vertices in the range set. Note that, according to this 
definition, 
if at 
times
$n$ and $n+1$ all the fitnesses are above $q$, then in the time interval
$[n,n+1]$ an avalanche of range $3$ and duration $1$ occurs, even though there 
were no 
fitnesses 
below $q$.

The fitnesses of the vertices are random variables with values
in $[0,1]$ and we update them according to the uniform
distribution on $[0,1]$. For computational reasons however, it
is convenient for the fitnesses to have values in $[0,\infty]$
and to update them according to the exponential distribution
with parameter~$1$, say.
In this new setup a threshold $b$ corresponds to the threshold
$q=1-e^{-b}$ in the old setup.

Suppose that a $b$-avalanche starts at time 0, with the origin
at the vertex $0$, so that
the vertex $0$ and its two neighbours are updated. We can now graphically
illustrate the $b$-avalanche on $\Lambda(N)\times \RR^+$ (space $\times$ 
fitness) as 
follows.
Look for the vertex with minimal fitness, and call this vertex $x$.
(Note that $x$  must be the vertex $0$ or one of its two neighbours.)
Suppose that the fitness of $x$ is equal to $s < b$. We then continue
updating according to the appropriate rules, and wait until all fitnesses
are above the threshold $s$. This in itself constitutes an $s$-avalanche,
starting at $x$. We denote by $\xi_N(x,s)$ the set of vertices involved in this
$s$-avalanche. In the graphical representation, we draw an arrow from the
space-fitness point $(x,s)$ to the space-fitness points $(i,s)$, for all
$i \in \xi_N(x,s)$.

After the $s$-avalanche has ended, the new fitnesses of all vertices
involved in this avalanche are i.i.d.\ and exponentially distributed
on $[s,\infty)$, due to the lack of memory property of the exponential
distribution. We can now look for the minimal fitness among all
vertices in $\xi_N(x,s)$. If this minimal fitness is above $b$,
then the $b$-avalanche has stopped. If this minimal fitness is equal to
$t$, where $s < t < b$, and is associated with the vertex $y$, say,
then we start, as before, a
$t$-avalanche with origin $y$. We continue updating until all fitnesses are 
above $t$.
If $\xi_N(y,t)$ denotes the set of vertices involved in this $t$-avalanche, then
we draw an arrow in the graphical representation from the space-fitness
point $(y,t)$ to all space-fitness points $(i,t)$, for $i \in \xi_N(y,t)$.
We continue in the obvious way. Under the assumption that all
avalanches are finite, this process will stop a.s. as soon as all
fitnesses are above $b$. The idea of avalanches which form a hierarchical 
structure of 
subavalanches 
is also mentioned in \cite{M}, in a slightly different context. 

This graphical representation, denoted by $GR_N$, is a random graph on the
space-fitness diagram $\Lambda(N) \times \RR^+$. We can describe it more
formally as follows.

Let $\{\Pi_k\}_{k\in\Lambda(N)}$ be a collection of independent homogeneous
Poisson processes. For each process $\Pi_k$ we perform the following procedure.
At the $j$-th arrival $\tau_{k,j}$ of $\Pi_k$,
we draw a pair $\bigg(\xit_N(k,\tau_{k,j}),\etat_N(\tau_{k,j})\bigg)$,
where $\xit_N(k,\tau_{k,j})$ is distributed as the range set, 
$\etat_N(\tau_{k,j})$ as the duration of a typical $\tau_{k,j}$-avalanche, 
with origin at $k$. We draw arrows in 
$\Lambda(N)\times\RR^+$ from $(k,\tau_{k,j})$ to
$(y,\tau_{k,j})$, for all $y\in \xit_N(k,\tau_{k,j})$.
For any $t_1<t_2$ we say that $(x,t_1)$ is {\em connected }to $(x,t_2)$
by a {\em time segment}. A {\em path} is a sequence $(x_0,s_0),\dots,(x_n,s_n)$ 
of 
points in $\Lambda(N) \times \RR^+$ such that every pair
$(x_j,s_j)$, $(x_{j+1},s_{j+1})$ is connected by either by
a time segment or an arrow.
For any $x,y\in \ZZ$, and $t_1\le t_2\in\RR$,
write $(x,t_1)\tot(y,t_2)$ in $GR_N$,
if there exists a path from $(x,t_1)$ to $(y,t_2)$.
See Figure~\ref{f4} for an illustration.
\fig{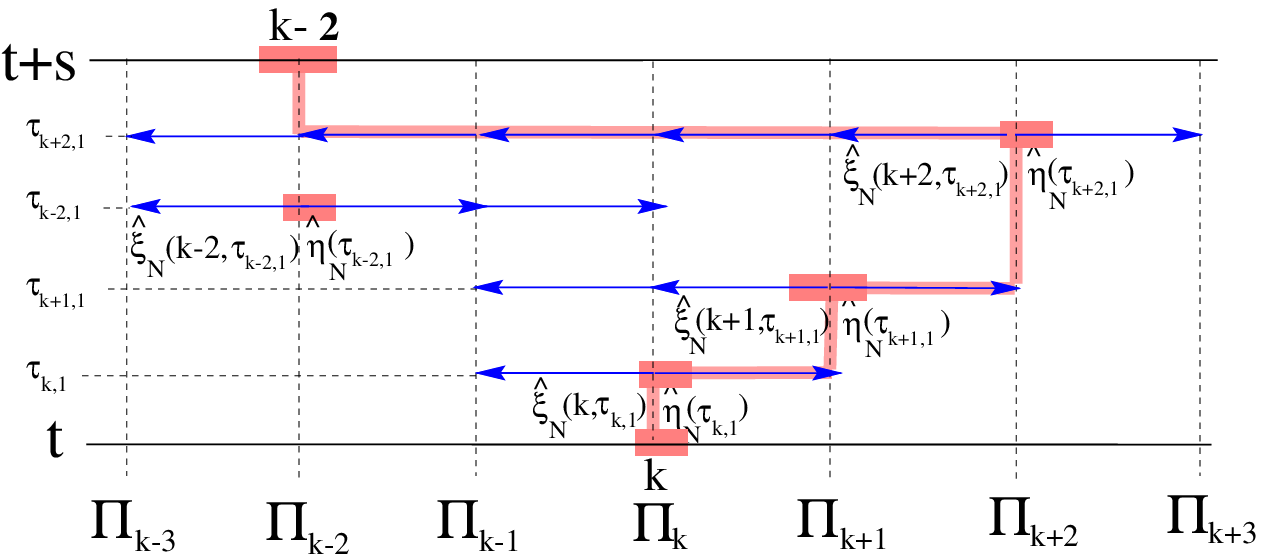}{The graphical representation $GR_N$,
where, for instance, $(k,t)\tot(k-2,t+s)$ in $GR_N$.}{f4}

For any $b>0$ the range set $\xi_N(0,b)$ of a $b$-avalanche with origin at $0$
consists of all vertices $x$ such that $(0,0)\tot(x,b)$
in $GR_N$, and the duration, denoted by $\eta_N(b)$, of this avalanche
is the sum of $\etat_N(\tau_{x,j})$ over all $\tau_{x,j}\le b$,
such that $(0,0)\tot(x,\tau_{x,j})$ in $GR_N$.

The graphical representation provides us with the following monotonicity 
properties. For any $A\subset\Lambda(N)$ and $t,s\ge0$
we denote by $\xi^{(A,t)}_N(s)$ the
random set of vertices $x\in\Lambda(N)$ such that there exists
$y=y(x)\in A$, and $(y,t)\tot(x,t+s)$ in $GR_N$.
Similarly, for any $A\subseteq\Lambda(N)$ and $t,s\ge0$,
we denote by $\eta_N^{(A,t)}(s)$ the sum of
$\etat_N(\tau_{x,j})$ over all $\tau_{x,j}\le t+s$,
such that there exists $y=y(x)\in A$,
and $(y,t)\tot(x,\tau_{x,i})$ in $GR_N$.
Then for any $A\subseteq B\subseteq\Lambda(N)$, $0\le s_1\le s_2$,
and $t\ge0$
\boe
\label{eq2}
\begin{array}{rl}
\xi^{(A,t)}_N(s_1)&\subseteq\xi^{(B,t)}_N(s_2),\\
\eta^{(A,t)}_N(s_1)&\le\eta^{(B,t)}_N(s_2).
\end{array}
\eoe
In particular, for $\xi_N(0,b)=\xi^{(\{-1,0,-N+1\},0)}_N(b)$
and $\eta_N(b)=\eta^{(\{-1,0,-N+1\},0)}_N(b)$, we have
\boe
\xi_N(0,b_1)\subseteq\xi_N(0,b_2),
\qquad\eta_N(b_1)\le\eta_N(b_2),\qquad\mbox{if }b_1\le b_2.
\eoe
The last inequality allows us to couple two copies $GR_N$ and $GR'_N$ of the 
graphical representation in such a way that for any $k\in\Lambda(N)$,
$\Pi_k(\cdot)$ restricted to $[0,b/2]$ is the same as $\Pi'_k(\cdot)$
restricted to $[b/2,b]$, and $\etat(\tau_{k,j})\le \etat '(\tau '_{k,j})$,
for $\tau_{k,j}\in[0,b/2]$ and $\tau '_{k,j}\in[b/2,b]$.
\fig{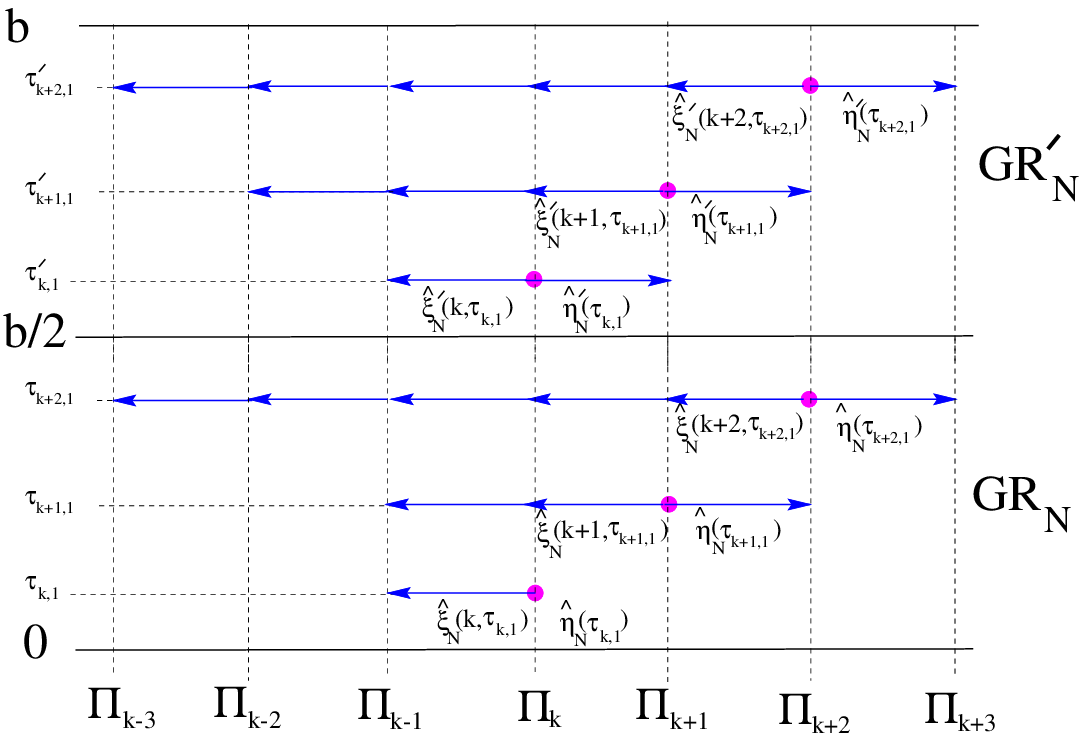}{The lower part of $GR_N$ and the upper part of $GR'_N$
coupled together.}{f8}
See Figure~\ref{f8} for an illustration of the coupled~$GR_N$ and~$GR'_N$.
Note that we do {\em not} claim that the two copies together yield 
a realisation of the evolution of the process from 0 to $b$.  
Nevertheless, this coupling gives us, for any $b>0$
$$
\eta^{(\Lambda(N),0)}_N(b/2)\le\eta'^{(\Lambda(N),b/2)}_N(b/2),$$
and hence
\boe
\label{eq26}
2 E\bigg(\eta^{(\Lambda(N),b/2)}_N(b/2)\bigg)\ge
  E\bigg(\eta^{(\Lambda(N),0)}_N(b)\bigg).
\eoe

\section{Proof of Theorem \ref{P1}}

An important step in the proof of Theorem \ref{P1} is the following lemma,
estimating the probability that an avalanche has range $N$, uniformly in $N$.
For any $b\in\RR_+$, define $P_N(b)$ as the probability that an avalanche
has range $N$.
\begin{lemma}
\label{L1}
If $b$ is large enough, then $P_N(b) \ge 1/2$, uniformly in $N$.
\end{lemma}
\begin{proof}{}
For any $x\in A\subseteq\Lambda(N)$, denote by ${\ell}(x,A)$ 
the {\em left corner} of $A$ with respect to $x$, 
$$
\ell\left(x,A\right)
=\min\bigg\{k\in(-\infty,x]\,\bigg|\,[k,x]\subseteq A\,\mbox{mod }N\bigg\}.$$
and write
$l_N(s):=\ell\left(0,\xi_N(0,s)\right)$ for the leftmost vertex
involved in an $s$-avalanche with the origin at 0. We will have proved 
the lemma if we show that $l_N$ is {\em explosive} in the sense 
that there exists a $0 < b_\infty<\infty$ such that for any
$0\le i\le i_{max}=\lceil\log_{3/2}(N-1)\rceil\in\NN$,
\boe
\label{eq4}
P\left(l_N(b_\infty)\le\max\Big\{-\left(\frac{3}{2}\right)^i,-N+1\Big\}\right)
\ge \frac{1}{2}+(\frac{1}{2})^{i+1},
\eoe
where the `max' actually only works at $i=i_{max}$.
Indeed, (\ref{eq4}) implies that $P_N(b_{\infty}) \geq \frac12$.
In order to achieve this, choose a constant $b_0\ge17$. Define a converging 
sequence of thresholds $b_1,b_2,b_3,\dots$ as follows:
$$
\begin{array}{rl}
b_i=&b_{i-1}+\left(\frac{3}{4}\right)^ib_0,\qquad i\ge1,\\
b_\infty=&\lim\limits_{i\to\infty}b_i=4b_0.
\end{array}$$
Observe that due to the monotonicity property~(\ref{eq2}),
it suffices to prove that for all $i\in[0,i_{max}]$
\boe
\label{eq5}
P\left(l_N(b_i)\le\max\Big\{-\left(\frac{3}{2}\right)^i,-N+1\Big\}\right)
\ge \frac{1}{2}+(\frac{1}{2})^{i+1}.
\eoe
We proceed by induction.
First note that
\boe
\label{eq6}
P\left(l_N(b_0)\le-1\right)=1.
\eoe
Next, suppose that~(\ref{eq5}) holds for some $i\in[0,i_{max}-1]$.
Observe that 
$$
\left\{
\begin{array}{c}
l_N(b_i)\le-\left(\frac{3}{2}\right)^i,\\
\exists 
x\in[-\left(\frac{3}{2}\right)^i,-\frac{1}{2}\left(\frac{3}{2}\right)^i]
\cap\Lambda(N),\;\exists \tau_{x,j}\in\Pi_x\cap[b_i,b_{i+1}),\\
\mbox{ such that }\;
\ell\left(x,\xit_N(x,\tau_{k,j})\right)\le x-\left(\frac{3}{2}\right)^i
\end{array}\right\}$$
implies
$$
\left\{
\begin{array}{c}
l_N(b_{i+1})\le\max\left\{-\left(\frac{3}{2}\right)^{i+1},-N+1\right\}
\end{array}\right\}.$$
\fig{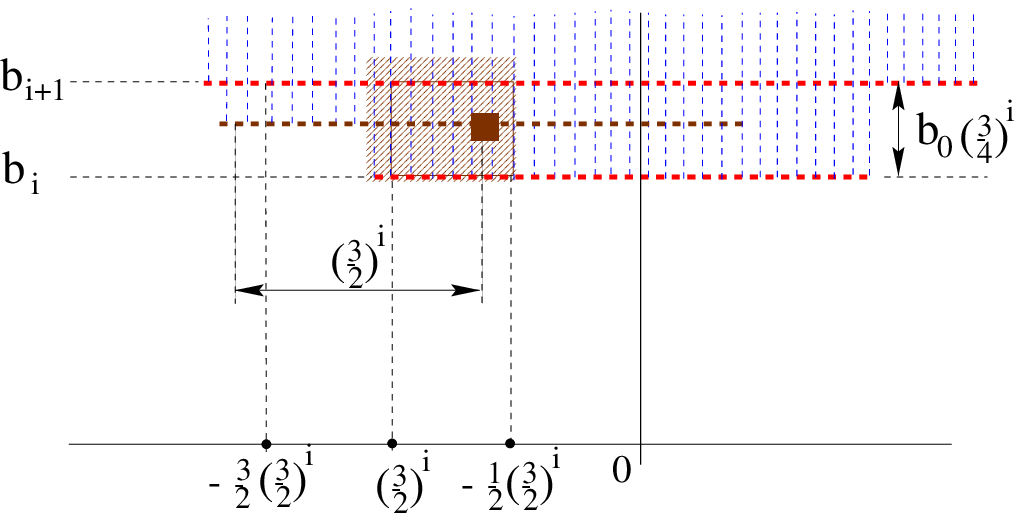}{Illustration of the induction step.}{f5}
See Figure~\ref{f5} for an illustration. Hence to finish the inductive step,
it suffices to show that for all $i\in[0,i_{max}-1]$,
\boe
\label{eq9}
P\left(
\begin{array}{c}
\forall 
x\in[-\left(\frac{3}{2}\right)^i,-\frac{1}{2}\left(\frac{3}{2}\right)^i]
\cap\Lambda(N),\;\forall \tau_{x,j}\in\Pi_x\cap[b_i,b_{i+1}),\\
\ell\left(x,\xit_N(x,\tau_{k,j})\right)> x-\left(\frac{3}{2}\right)^i
\end{array}\right)
\le\left(\frac{1}{2}\right)^{i+2}.
\eoe
Since the $\ell\left(x,\xit_N(x,\tau_{k,j})\right)$'s
are independent and since (due to the monotonicity property) 
for any $x \in \Lambda(N)$ and $\tau_{x,j}\ge b_i$,
$$
\begin{array}{c}
P\left(\ell(x,\xit_N(x,\tau_{x,j}))>x-\left(\frac{3}{2}\right)^i\right)
\le P\left(l_N(b_i)>-\left(\frac{3}{2}\right)^i\right)\le \frac{1}{2},
\end{array}$$
the points
\boe
\label{eq10}
\bigcup\limits_
{x\in[-\left(\frac{3}{2}\right)^i,-\frac{1}{2}\left(\frac{3}{2}\right)^i]
\cap\Lambda(N)}\left\{\tau_{x,j}\in\Pi_x\,;
\,\ell\left(x,\xit_N(x,\tau_{x,j})\right)
\le x -\left(\frac{3}{2}\right)^i\right\}
\eoe
constitute a thinning of the Poisson process $\bigcup\limits_
{x\in[-\left(\frac{3}{2}\right)^i,-\frac{1}{2}\left(\frac{3}{2}\right)^i]
\cap\Lambda(N)}\Pi_x$ with deleting probability at most $1/2$.
Thus the points in ~(\ref{eq10}) contain a Poisson process of intensity 
at least
$$
\begin{array}{c}
\frac{1}{2}\cdot\Big(\mbox{intensity of }\bigcup\limits_
{x\in[-\left(\frac{3}{2}\right)^i,-\frac{1}{2}\left(\frac{3}{2}\right)^i]
\cap\Lambda(N)}\Pi_x\Big)\\
\ge\frac{1}{2}\left(\lfloor \left(\frac{3}{2}\right)^i\rfloor
-\lfloor\frac{1}{2}\left(\frac{3}{2}\right)^i\rfloor\right)
\ge\frac{1}{2}\cdot\frac{1}{3}\left(\frac{3}{2}\right)^i.
\end{array}$$
Observe that the event in~(\ref{eq9})
implies that the process~(\ref{eq10}) has no arrivals between time
$b_{i}$ and $b_{i+1}$, a time interval of length $b_0(\frac{3}{4})^i$, 
and hence it has probability at most
$$
\begin{array}{c}
\exp\{-\frac{1}{6}\left(\frac{3}{2}\right)^i b_0\left(\frac{3}{4}\right)^i\}
=\exp\{-\frac{b_0}{6}\left(\frac{9}{8}\right)^{i}\}
\le \left(\frac{1}{2}\right)^{i+2},\qquad i\in\NN,
\end{array}
$$
since $b_0\ge17$.
So we have~(\ref{eq9}) and the proof is complete.
\end{proof}

\begin{proof}{ of Theorem~\ref{P1}}
Suppose we start from an i.i.d.\ uniform
distribution above the threshold $q<1$, which we assume 
however is so close to one that in the model where we update fitnesses 
according to the exponential distribution, we would have $P_N(b/2) >1/2$, 
uniformly in $N$. 
(Recall that $b$ and $q$ are related via $q=1-e^{-b}$.)
We will define the dynamics via the following 
independent sequences of i.i.d. random variables. Fix some $q'\in(q,1)$.
Let, for $i=1,2,3$, ${\cal U}^{i}=(U^{i}_j)_{j\in\NN}$ be a sequence of 
i.i.d.\ random variables uniformly distributed on $[q',1]$.
We will use ${\cal U}^{i}$ to construct the dynamics above threshold $q'$.
Let, for $i=1,2,3$, ${\cal V}^{i}=(V^{i}_j)_{j\in\NN}$ be a sequence of 
i.i.d.\ random variables uniformly distributed on $[0,q']$.
We will use ${\cal V}^{i}$ to construct the dynamics below threshold $q'$.
Let, for $i=1,2,3$, ${\cal S}^{i}=(S^{i}_j)_{j\in\NN}$ be a sequence of 
i.i.d.\ Bernoulli distributed random variables taking the value $1$
with probability $1-q'$ and the value $0$ with probability $q'$.
We will use ${\cal S}^{i}$ to choose between ${\cal U}^{i}$ and 
${\cal V}^{i}$, and store the result in the sequence 
${\cal G}^{i}=(G^{i}_j)_{j\in\NN}$, $i=1,2,3$:
$$
G^{i}_j=\left\{
\begin{array}{rl}
U^{i}_j,&\mbox{ if }S^{i}_j=1,\\
V^{i}_j,&\mbox{ if }S^{i}_j=0.
\end{array}\right.\qquad j\in\NN.$$
It is clear that for $i=1,2,3$, the sequence ${\cal G}^{i}$ consists of i.i.d.\
random variables uniformly distributed on $[0,1]$. 
We will use the beginning of the sequence ${\cal G}^{2}$ (this choice is 
arbitrary)
to assign the initial fitnesses to the vertices.
At time $n\in[0,N-1]$,
we use the random variable $G^2_n$ to assign an
initial fitness to the vertex with number $-n$. 
Now we are ready to define the dynamics.
At every time $n\ge N$
we choose the vertex $k$, say, with minimal fitness at
time $n-1$, and we assign to $(k-1,k,k+1)(\mbox{ mod }N)$
the triple $(G^1_n,G^2_n,G^3_n)$. 

Let $f_0(n)$ denote the fitness of a fixed observation vertex at time $n$.
Let $j_n$ be the moment that this vertex received its current fitness, i.e.,
$$
j_n=\min\Big\{j\le n\,|\,f_0(j)=f_0(n)\Big\}.$$
Let $i(n)$ be the number of the sequence providing this value, i.e.
$$
f_0(n)=G^{i(n)}_{j_n}.$$
Observe that (since $j_{j_n}=j_n$) $i(n)=i(j_n)$.
We say that $n$ is $q'${\em-good}, if during $[j_n,n]$
the minimal fitness of all vertices is always less than $q'$. 
Then, for any time $n\in\NN$ and any $q''\in(q',1)$, we have
\boe
\label{eq11}
\begin{array}{rl}
P\Big(f_0(n)<q''\Big)
&=\sumu\limits_{j=0}^{n-1}P\Big(f_0(n)<q'',\,j_n=j\Big)\\
&=\sumu\limits_{i=1}^{3}\sumu\limits_{j=0}^{n-1}
P\Big(G^i_j<q'',\,j_n=j,\,i(j)=i\Big)\\
&\ge\sumu\limits_{i=1}^{3}\sumu\limits_{j=0}^{n-1}
P\Big(G^i_j<q'',\,j_n=j,\,i(j)=i,\,n\mbox{ is }q'-\mbox{good}\Big)\\
&\ge\sumu\limits_{i=1}^{3}\sumu\limits_{j=0}^{n-1}
P\Big(U^i_j<q'',\,j_n=j,\,i(j)=i,\,n\mbox{ is }q'-\mbox{good}\Big).
\end{array}
\eoe
Now observe that for any $j\in[0,n]$, the 
event~$\Big\{j_n=j,\,i(j)=i,\,n\mbox{ is }q'-\mbox{good}\Big\}$
is measurable with respect to 
$$
\sigma\Big\{U^i_{j'},V^i_{j'},S^i_{j'},\,
S^i_{j''},V^i_{j''},\,0\le j'<j,\,j\le j''\le n,\,i=1,2,3\Big\}.$$
Hence, for any $j\in[0,n]$, the events $\Big\{U^i_j<q''\Big\}$ 
and~$\Big\{j_n=j,\,i(j)=i,\,n\mbox{ is }q'-\mbox{good}\Big\}$ 
are independent, and we can continue the estimate~(\ref{eq11}) as
\boe
\label{eq111}
\begin{array}{l}
=\sumu\limits_{i=1}^{3}\sumu\limits_{j=0}^{n-1}
P\Big(U^i_j<q''\Big)
P\Big(j_n=j,\,i(j)=i,\,n\mbox{ is }q'-\mbox{good}\Big)\\
=\frac{q''-q'}{1-q'}\sumu\limits_{i=1}^3\sumu\limits_{j=0}^{n-1}
P\Big(j_n=j,\,i(j)=i,\,n\mbox{ is }q'-\mbox{good}\Big)\\
=\frac{q''-q'}{1-q'}P\Big(n\mbox{ is }q'-\mbox{good}\Big).
\end{array}
\eoe
It remains to estimate the probability that 
$n$ is $q'$-good from below, uniformly in $n$ and~$N$.

Define a sequence $(\tau^N_j)_{j\in\NN}$
of stopping times, with respect to the natural filtration, as follows:
$\tau_0=0$, and for any $j\in\NN$, $\tau_{j+1}$
is the end of the first $q$-avalanche of range $N$
after $\tau_j$. For any $j\in\NN$, we call the time interval
$I^N_j=[\tau_j,\tau_{j+1})$ the $j$-th
{\em period}. It is clear that at
every $\tau^N_j$ the fitnesses are i.i.d.\ and
uniformly distributed above the threshold $q$.
Thus the period lengths are i.i.d.\ random variables.
For any time $n$, we denote by $\tau_{j(n)}$  
the maximal $\tau_j$ such that $\tau_j\le n$, i.e.,
$n\in [\tau_{j(n)},\tau_{j(n)+1})$, 
and we say that $n$ is $q'$-{\em nice}, if during $[\tau_{j(n)},n]$
the minimal fitness is always less than $q'$.
For any $n\in\NN$, if $n$ is $q'$-nice then $n$ is $q'$-good.
Indeed, suppose that $n$ is $q'$-nice. If $j_n\ge \tau_{j(n)}$,
then $n$ is clearly $q'$-good. Suppose $j_n< \tau_{j(n)}$.
Since during the $q$-avalanche of range $N$ of the previous period,
every vertex has been updated, $j_n$ belongs to this $q$-avalanche,
and hence the minimal fitness at the time interval $[j_n,\tau_{j(n)})$ 
is always less than $q<q'$, and $n$ is $q'$-good.
Thus it suffices to show that the probability that time 
$n$ is $q'$-nice, is bounded away from zero, uniformly in $N$ and $n$.

A period can be decomposed into two
parts: the duration of the avalanche of range $N$, and the waiting time until
this avalanche. We denote by $W_N$ a typical waiting time before
the avalanche of range $N$, and by $A_N$ the duration of this avalanche.
During a $q$-avalanche, the minimal fitness is always at most $q<q'$.
Hence, if in the $i$-th period, the waiting time $W_N$ satisfies $W_N=0$,
and in addition there is at least one vertex  at time $\tau_i^N$ with
fitness between $q$ and $q'$, then any time $n$ within the $i$-th period is
$q'$-nice.
The event that there is such a vertex with fitness between $q$ and $q'$
is independent of $W_N$ and $A_N$ associated to that period, and
has probability
$$
p_1=1-\left(1-\frac{q'-q}{1-q}\right)^N\ge \frac{q'-q}{1-q}>0,
\qquad\mbox{ uniformly in }N.$$
Hence, in the stationary regime with $N$ vertices, we can write 
(using alternating renewal process theory),
\begin{eqnarray*}
P(n \mbox{ is }q'\mbox{-nice})
&\geq& p_1 P(n \mbox{ is in a period with }W_N=0)\\
&\to& p_1 P\big(W_N=0\big)\frac{E(A_N)}{E(W_N)+E(A_N)},
\end{eqnarray*}
for $n \to \infty$. At this point, we switch from $q$ to $b$, 
since we use results from the previous section. 

Since $P\big(W_N=0\big)=P_N(b)\ge P_N(b/2)\ge 1/2$, uniformly in $N$,
it suffices to prove that there exists a constant $0<c(b)<\infty$,
independent of~$N$, such that
\boe
\label{eq25}
E(W_N)\le c(b) E(A_N).
\eoe
Denote by $Y_N$ the number of $b$-avalanches preceding the $b$-avalanche
of range $N$. Every avalanche has range $N$ with probability
$P_N(b)$, independently of all other avalanches.
 Hence $Y_N+1$ has 
a geometrical distribution with parameter $P_N(b)$, and we have
$$
E(Y_N)=\frac{1}{P_N(b)}-1.$$
Let $(Z_i^N)_{i\in\NN}$ be an i.i.d.\ sequence of random variables
distributed as the duration of a typical $b$-avalanche, conditioned 
on its range being smaller than $N$.
Then we can  use $Y_N$ of those avalanches to obtain $W_N$, i.e.
$$
W_N=Z_1^N + \cdots + Z_{Y_N}^N.$$
In words, at the beginning of a new avalanche, we first decide 
(with the correct probability) whether 
or not the avalanche has range $N$. 
If not, we choose one, conditioned on its range being smaller 
than $N$, and the resulting duration is the next $Z_i^N$.
Since due to the construction, $Y_N$ is independent of the sequence 
$(Z_i^N)_{i\in\NN}$, we have that
\boe
\label{eq12}
\begin{array}{rl}
E(W_N)&=E(Z_1^N + \cdots + Z_{Y_N}^N)\\
&= E(Y_N)E(Z_1^N)=\left(\frac{1}{P_N(b)}-1\right)E(Z_1^N).
\end{array}
\eoe
We will now estimate $E(Z_1^N)$ from above and $E(A_N)$ from below
(recall that the pair $(\xi_N(b), \eta_N(b))$ represents 
the range set and duration of a $b$-avalanche)
$$
\begin{array}{rl}
E(Z_1^N)&=E\bigg(\eta_N(b)\,\bigg|\,|\xi_N(b)|\le N-1\bigg)\\
&= \sumu\limits_{k=0}^{\infty} 
\frac{k P(\eta_N(b)=k, |\xi_N(b)| \leq N-1)}{P(|\xi_N(b)|\leq N-1)}\\
&\le \sumu\limits_{k=0}^{\infty} \frac{kP(\eta_N(b)=k)}{P(|\xi_N(b)| \leq 
N-1)}\\
&\le E\bigg(\eta_N(b)\bigg)/P\bigg(|\xi_N(b)|\le N-1\bigg)\\
&= \frac{1}{1-P_N(b)}E\bigg(\eta_N(b)\bigg).\\
E(A_N)&=E\bigg(\eta_N(b)\,\bigg|\,|\xi_N(b)|=N\bigg)\\
&\ge E\bigg(\eta_N(b)\indic\{|\xi_N(b)|=N\}\bigg)\\
&\ge E\bigg(\eta_N(b)\indic\{|\xi_N(b/2)|=N\}\bigg)\\
&\ge E\bigg(\eta^{(\Lambda(N),b/2)}_N(b/2)\indic\{|\xi_N(b/2)|=N\}\bigg),
\end{array}$$
and, since $\eta^{(\Lambda(N),b/2)}_N(b/2)$ and $\indic\{|\xi_N(b/2)|=N\}$
are independent, this is equal to
$$
E\bigg(\eta^{(\Lambda(N),b/2)}_N(b/2)\bigg)
P\bigg(|\xi_N(b/2)|=N\bigg)\ge \frac{1}{2} E\bigg(\eta_N(b)\bigg)P_N(b/2),$$
where the last inequality follows from~(\ref{eq26}).
Combining the estimates of $E(Z_1^N)$ and $E(A_N)$, we have
$$
E(Z_1^N)\le \frac{1}{1-P_N(b)}\frac{2}{P_N(b/2)} E(A_N).$$
Since $P_N(b)\ge P_N(b/2) \geq 1/2$, the above inequality
together with~(\ref{eq12}) gives us
$$
\begin{array}{rl}
E(W_N)
&\le 
\left(\frac{1}{P_N(b)}-1\right)\frac{1}{1-P_N(b)}\cdot\frac{2}{P_N(b/2)}E(A_N)\\
&\le \frac{2}{P_N(b)P_N(b/2)}E(A_N)\le8E(A_N).
\end{array}$$
Thus we have~(\ref{eq25}) and the theorem.
\end{proof}

\section{Extension to general graphs}
\label{G}
Let $G$ be a finite connected graph.
One can define on $G$ a Bak-Sneppen process in the following way. 
We call two vertices neighbours if they are connected by a bond 
in $G$. Every vertex of $G$ accomodates a random variable (the fitness), 
with value in $[0,1]$. At the initial moment, all the fitnesses are i.i.d.\ 
and uniformly distributed on $[0,1]$. Every discrete time step we choose 
a vertex with minimal fitness, and replace it, together with the fitnesses 
of all its neighbours, 
by new independent fitnesses, uniformly distributed in $[0,1]$. We will
give bounds for the mean of the fitness in the stationary regime.
These bounds will depend on the local geometrical structure of $G$, 
but will be independent of the number of vertices in $G$. 

In order to state our main result, here follows some notation.
Let $V_G$ denote the set of vertices of $G$.
For any two vertices $x,y\in V_G$, we denote by $\rho_G(x,y)$ the distance
between them, i.e.\ the number of bonds in the shortest path between 
$x$ and $y$. Then for any vertex $x\in V_G$ and $k\in\NN$, 
we can define the ball
$$
B^x_G(k)=\{y\in V_G\,|\,\rho(y,x)\le k\},$$
and the sphere
$$
S^x_G(k)=\{y\in V_G\,|\,\rho_G(y,x)= k\}.$$
Observe that since $G$ is finite, we have, for sufficiently large $k=k(G)$, 
that $B^x_G(k)=V_G$ and $S^x_G(k)=\emptyset$, for any $x\in V_G$.

For any $k\in\NN$ denote by $m_G(k)$ the number 
of vertices in the smallest ball of radius $k$ and denote by $M_G(k)$
the number of vertices in the largest sphere of radius $k$, i.e.
$$
\begin{array}{rl}
m_G(k)&=\min\limits_{x\in V_G}|B^x_G(k)|,\\
M_G(k)&=\max\limits_{x\in V_G}|S^x_G(k)|.
\end{array}$$
Consider an increasing sequence of radii 
\boe
\label{eqg15}
\begin{array}{c}
r_i=\left\lfloor\left(\frac{4}{3}\right)^{i+2}\right\rfloor,\qquad i\in\NN.
\end{array}
\eoe

For any vertex $x$ of $G$ we denote by $F^x_G$ the distribution function 
of the fitness at $x$ in the stationary regime. In the following two theorems
we will establish an analogue of the phase-transition result for 
an infinite collection of finite connected graphs.

\begin{theorem}
\label{ThG}
Let ${\cal G}$ be an infinite collection of finite connected graphs 
such that for some $60\le b<\infty$,
\boe
\label{eqThG}
\frac{b}{4}\Big(\frac{4}{5}\Big)^i m_G(\lfloor r_i/3\rfloor)
\ge\log\Big(M_G(\lceil 2r_i/3\rceil)\Big),
\qquad\mbox{uniformly in }G\in{\cal G}, i\in\NN.
\eoe
Then there exists $c_b>0$ and $q_b\in(0,1)$ such that 
$$
F^x_G(q_b)>c_b,\qquad\mbox{uniformly in }G\in{\cal G},x\in G.$$
\end{theorem}

\begin{theorem}
\label{ThG1}
Let ${\cal G}$ be an infinite collection of finite connected graphs of 
uniformly bounded degree, i.e.\ there exists a constant $K\in\NN$
such that 
\boe
\label{eqThG1}
\max\limits_{x\in V_G}|B^x_G(1)|<K,
\qquad\mbox{uniformly in }G\in{\cal G}.
\eoe
Then for any sequence $(G_n)_{n\in\NN}\subset {\cal G}$ such that
$$
|V_{G_n}|\to\infty,\qquad\mbox{as }n\to\infty,$$
we have 
$$
\lim\limits_{n\to\infty}\left(\max_{x \in V_{G_n}} F^x_{G_n}(1/K)\right)=0.$$
\end{theorem}
The proof uses a standard branching process argument and is omitted.

\subsection{Examples}

\medskip\noindent
{\bf 1. The original Bak-Sneppen model on the circle.}
Here ${\cal G}=(P_n)_{n\in\NN},$
where $P_n$ is the regular polygon with $n$ vertices. 
Observe that for any $k\ge n$, we have $M_{P_n}(\lceil 2k/3\rceil)=0$.
For any $k<n$, we have 
$$
\begin{array}{rl}
m_{P_n}(\lfloor k/3\rfloor)\ge 2k/3,
&\qquad M_{P_n}(\lceil 2k/3\rceil)\le 2.
\end{array}$$
Then condition~(\ref{eqThG}) holds for $b\ge60$,
and condition~(\ref{eqThG1}) holds for $K=3$.

\medskip\noindent
{\bf 2. The multi-dimensional Bak-Sneppen model.}
Let $d\ge 2$ be the dimension. We consider 
${\cal G}=(G_n)_{n\in\NN}$, where for every $n\in\NN$ 
the set of vertices $V_{G_n}=\{1,\dots,n\}^d$, and 
with the usual nearest neighbour structure with periodic boundary conditions.
Observe that for any $k\ge n$ we have $M_{P_n}(\lceil 2k/3\rceil)=0$.
For any $k<n$, we have 
$$
\begin{array}{rl}
m_{P_n}(\lfloor k/3\rfloor)\sim c_1 k^d,
&\qquad M_{P_n}(\lceil 2k/3\rceil) \sim c_2 k^{d-1}.
\end{array}$$
Property~(\ref{eqThG}) holds 
for $b\ge30d$, and condition~(\ref{eqThG1}) holds 
for $K=2d+1$.

\medskip\noindent
{\bf 3. The Bak-Sneppen model on a tree.}
Choose $d\ge 2$ and consider ${\cal G}=(T_d(n))_{n\in\NN},$
where $T_d(n)$ is the regular $d$-ary tree with $d$ offsprings
at each vertex and $n$ generations. 
Observe that for any $k>3n$ we have $M_{P_n}(\lceil 2k/3\rceil)=0$.
For any $k\le3n$, we have 
$$
\begin{array}{rl}
m_{P_n}(\lfloor k/3\rfloor) \sim c_3 d^{k/6},
&\qquad M_{P_n}(\lceil 2k/3\rceil)\sim c_4 d^{2k/3}.
\end{array}$$
One can check that property~(\ref{eqThG}) holds for $b\ge30d$,
and that condition~(\ref{eqThG1}) holds for $K=d+2$.

\section{Proof of Theorem~\ref{ThG}}
The proof of Theorem~\ref{ThG} essentially follows 
the proof of Theorem~\ref{P1}. 
For any finite $G$, one can associate a graphical
representation $GR_G$. The construction of this graphical representation
is essentially the same as before.
Let $\{\Pi_k\}_{k\in V_G}$ be a collection of
independent homogeneous Poisson processes. 
For each process $\Pi_k$ we perform the following procedure.
At the $j$-th arrival $\tau_{k,j}$ of $\Pi_k$,
we draw a pair $\bigg(\xit_G(k,\tau_{k,j}),\etat_G(k,\tau_{k,j})\bigg)$,
where $\xit_G(k,\tau_{k,j})$ is distributed as the range set, 
$\etat_G(k,\tau_{k,j})$ as the duration of a typical $\tau_{k,j}$-avalanche
with origin at $k$. 
We draw arrows in 
$V_G\times\RR^+$ from $(k,\tau_{k,j})$ to
$(y,\tau_{k,j})$, for all $y\in \xit_G(k,\tau_{k,j})$.
As before, we can define, 
for any $A\subseteq V_G$, $0\le t,s<\infty$,
the processes $\xi^{(A,t)}_G(s)$ and $\eta^{(A,t)}_G(s)$,
such that the monotonicity properties hold
\boe
\label{eqg2}
\begin{array}{rl}
\xi^{(A,t)}_G(s_1)&\subseteq\xi^{(B,t)}_N(s_2),\\
\eta^{(A,t)}_G(s_1)&\le\eta^{(B,t)}_N(s_2),\qquad 0\le s_1\le s_2.
\end{array}
\eoe
The range set $\xi_G(x,b)$ and the duration $\eta_G(x,b)$ 
of a $b$-avalanche with origin at $x\in V_G$ can be written as 
$\xi_G(x,b)=\xi^{(B^x_G(1),0)}_G(b)$
and $\eta_N(x,b)=\eta^{(B^x_G(1),0)}_G(b)$.

The only place where we previously used the geometrical structure of $G$
is Lemma~\ref{L1} and the related definitions.
We will give a new lemma for a collection of finite connected graphs.

For any vertex $x$ of $G$ and any $q>0$
define $P^x_G(q)$ as the probability that updating 
$x$ and the neighbours in the configuration 
with all fitnesses above $q$ results in a $q$-avalanche
of range $|V_G|$. 

\begin{lemma}
\label{Lg1}
Let $G$ be a finite connected graph
such that for some $60\le b_0<\infty$,
$$
\begin{array}{c}
\frac{b_0}{4}\Big(\frac{4}{5}\Big)^i m_G(\lfloor 
r_i/3\rfloor)\ge\log\Big(M_G(\lceil2r_i/3\rceil)\Big),
\qquad\mbox{uniformly in }i\in\NN.
\end{array}$$
Then there exists $q_{\infty}(b_0)\in(0,1)$, depending only on $b_0$ 
such that for any $q>q_{\infty}(b_0)$
$$
P^x_G(q)>1/2,\qquad\mbox{uniformly in }x\in G.$$
\end{lemma}
\begin{proof}{}
As in the proof of Lemma~\ref{L1} we will work with
fitnesses defined on $[0,\infty)$, and we will update 
them according to the exponential distribution with parameter 1, say.
Recall that in the new setup a threshold $b$ corresponds to the threshold
$q=1-e^{-b}$ in the old setup.

We will have proved the lemma
if we show that for any $x\in G$, the process $\xi_G(x,t)$ 
is {\em explosive} in the sense that 
there exists $b_\infty\in(0,\infty)$, depending only on $b_0$,
such that for any $i\in\NN$ (essentially for 
$i\le i_{max}=\max\limits_{x,y}\rho(y,z)$, because 
for $i>i_{max}$ we have $B^x_G(r_i)\equiv V_G$)
\boe
\label{eqg4}
P\Big(B^x_G(r_i)\subseteq \xi_G(x,b_\infty)\Big)
\ge \frac{1}{2}+(\frac{1}{2})^{i+1}.
\eoe
Indeed, (\ref{eqg4}) implies that $P^x_G(b_{\infty}) \geq \frac12$.
In order to achieve this, choose a constant $b_0$ satisfying 
the condition of the lemma. Define a 
converging sequence of thresholds $b_1,b_2,b_3,\dots$ 
as follows:
$$
\begin{array}{rl}
b_i=&b_{i-1}+\left(\frac{4}{5}\right)^ib_0,\qquad i\ge1,\\
b_\infty=&\lim\limits_{i\to\infty}b_i=5b_0.
\end{array}$$
Observe that due to the monotonicity property~(\ref{eqg2}),
it suffices to prove that for all $i\in\NN$, $x\in V_G$
\boe
\label{eqg5}
P\left(B^x_G(r_i)\subseteq \xi_G(x,b_i)\right)
\ge \frac{1}{2}+(\frac{1}{2})^{i+1}.
\eoe
We proceed by induction.
First note that for any $x\in V_G$
\boe
\label{eqg6}
P\left(B^x_G(1)\subseteq \xi_G(x,b_0)\right)=1.
\eoe
Next, suppose that~(\ref{eqg5}) holds for some $i\in\NN$ and 
all $x\in V_G$.
Observe that 
\boe
\label{eqg7}
\left\{
\begin{array}{c}
B^x_G(r_i)\subseteq \xi_G(x,b_i),\\
\forall z\in S^x_G(\lceil2r_i/3\rceil),\,
\exists y\in B^z_G(\lfloor r_i/3\rfloor),\,
\exists \tau_{y,j}\in\Pi_y\cap[b_i,b_{i+1}),\\
\mbox{ such that }\;
B^y_G(r_i)\subseteq \xit_G(y,\tau_{y,j})
\end{array}\right\}
\eoe
implies
$$
\left\{
\begin{array}{c}
B^x_G(r_{i+1})\subseteq \xi_G(x,b_{i+1})
\end{array}\right\}.$$
Indeed, if $a\in B^x_G(r_{i+1})\setminus B^x_G(r_i)$ then 
there exists $z=z(a)\in S^x(\lceil2r_i/3\rceil)$
such that $\rho(a,z)+\rho(z,x)\le r_{i+1}$, 
and hence $\rho(a,z)\le \lceil2r_i/3\rceil$. Then if 
there exists $\tau\in\bigcup\limits_
{y\in B^z_G(\lfloor r_i/3\rfloor)}
\Pi_y\cap[b_i,b_{i+1})$ such that 
$B^y_G(r_i)\subseteq \xit_G(y,\tau)$ 
then $\rho(a,y)\le\rho(a,z)+\rho(z,y)\le r_i$,
and hence $a\in \xit_G(y,\tau)$. See
Figure~\ref{f1} for an illustration of~(\ref{eqg7}).
\fig{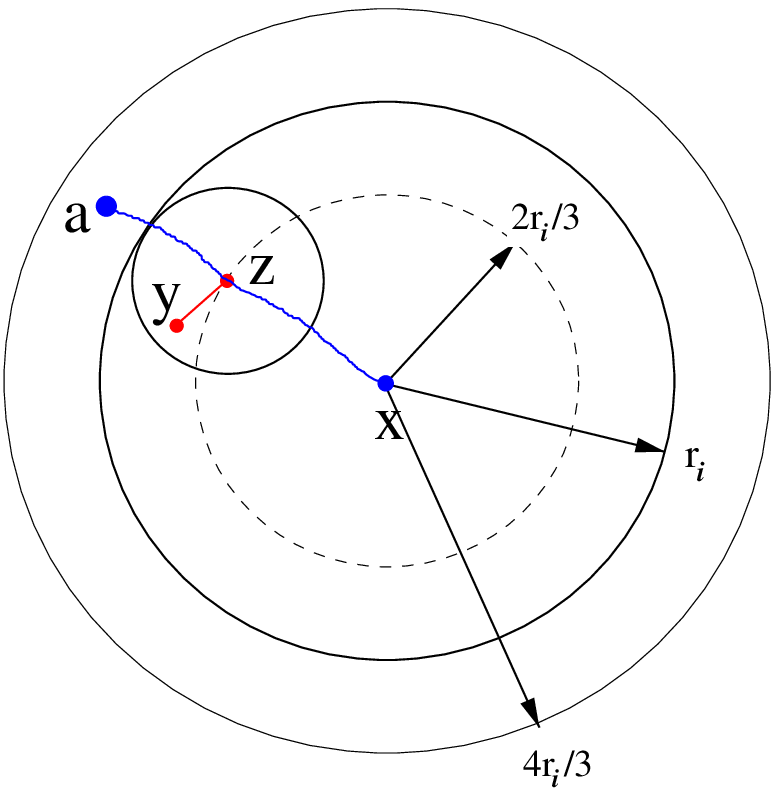}{Illustration of~(\ref{eqg7}).}{f1}
Hence to finish the inductive step,
it suffices to show that, uniformly in $i\in\NN$,
\boe
\label{eqg9}
P\left(
\begin{array}{c}
\exists z\in S^x_G(\lceil2r_i/3\rceil)\mbox{ such that }\\
\forall y\in B^z_G(\lfloor r_i/3\rfloor),\,
\forall \tau_{y,j}\in\Pi_y\cap[b_i,b_{i+1}),\\
B^y_G(r_i)\not\subseteq \xit_G(y,\tau_{y,j})
\end{array}\right)
\le\left(\frac{1}{2}\right)^{i+2}.
\eoe
Since the events $B^y_G(r_i)\subseteq \xit_G(y,\tau_{y,j})$
are independent and since (due to the monotonicity property) 
for any $y\in V_G$ and $\tau_{y,j}\ge b_i$,
$$
\begin{array}{c}
P\left(B^y_G(r_i)\subseteq \xit_G(y,\tau_{y,j})\right)
\le P\left(B^y_G(r_i)\subseteq \xit_G(y,b_i)\right)\le \frac{1}{2},
\end{array}$$
for any $z\in S^x_G(\lceil2r_i/3\rceil)$ the points
\boe
\label{eqg10}
\bigcup\limits_
{y\in B^z_G(\lfloor r_i/3\rfloor)}
\left\{\tau_{y,j}\in\Pi_y\,
; \,B^y_G(r_i)\subseteq \xit_G(y,\tau_{y,j})\right\}
\eoe
constitute a thinning of the Poisson process $\bigcup\limits_
{y\in B^z_G(\lfloor r_i/3\rfloor)}\Pi_y$
with deleting probability at most $1/2$.
Thus the points in ~(\ref{eqg10}) contain a Poisson process 
of intensity at least
$$
\frac{1}{2}\cdot\Big(\mbox{intensity of }\bigcup\limits_
{y\in B^z_G(\lfloor r_i/3\rfloor)}\Pi_y\Big)
\ge\frac{1}{2}\cdot m_G(\lfloor r_i/3\rfloor).
$$
Observe that the event in~(\ref{eqg9})
implies that for some $z\in S^x_G(\lceil2r_i/3\rceil)$,
the process~(\ref{eqg10}) has no arrivals between time
$b_{i}$ and $b_{i+1}$, a time interval of length $b_0(\frac{4}{5})^i$.
The last event has probability at most
$$
\begin{array}{c}
\exp\left\{-\frac{1}{2}
m_G(\lfloor r_i/3\rfloor)
b_0\left(\frac{4}{5}\right)^i\right\},
\end{array}$$
uniformly in $z\in S^x_G(\lceil2r_i/3\rceil)$. 
Thus we can estimate the probability in~(\ref{eqg9}) by 
$$
\begin{array}{c}
M_G(\lceil2r_i/3\rceil)\exp\{-\frac{b_0}{2}
m_G(\lfloor r_i/3\rfloor)\left(\frac{4}{5}\right)^i\}.
\end{array}$$
Split the above expression in two terms
$$
\begin{array}{c}
\left(M_G(\lceil2r_i/3\rceil)
\exp\left\{-\frac{b_0}{4}
m_G(\lfloor r_i/3\rfloor)\left(\frac{4}{5}\right)^i\right\}\right)
\exp\left\{-\frac{b_0}{4}
m_G(\lfloor r_i/3\rfloor)\left(\frac{4}{5}\right)^i\right\}.
\end{array}$$
The first term is less than or equal to $1$, under the conditions of the lemma. 
For
the second term, we write 
$$
\begin{array}{c}
\exp\left\{-\frac{b_0}{4}
m_G\left(\lfloor r_i/3\rfloor\right)\left(\frac{4}{5}\right)^i\right\}
\le\exp\left\{-\frac{b_0}{4}
\left(1+\lfloor r_i/3\rfloor\right)\left(\frac{4}{5}\right)^i\right\}\\
\le\exp\left\{-\frac{b_0}{4}
\left(\frac{1}{3}\left(\frac{4}{3}\right)^i\right)
\left(\frac{4}{5}\right)^i\right\}
\le\exp\left\{-\frac{b_0}{12}\left(\frac{16}{15}\right)^i\right\}
\le\left(\frac{1}{2}\right)^{i+2},\qquad i\in\NN,
\end{array}$$
since $b_0\ge60$.
So we have~(\ref{eqg9}) and the proof is complete.
\end{proof}

\begin{center}
{\bf Authors}
\end{center}

{\sc Ronald Meester}, Divisie Wiskunde,
Faculteit der Exacte Wetenschappen, Vrije Universiteit Amsterdam, 
de Boelelaan 1081a, 1081 HV Amsterdam, The Netherlands; e-mail:
{\tt rmeester@cs.vu.nl},

{\sc Dmitri Znamenski}, Divisie Wiskunde,
Faculteit der Exacte Wetenschappen, Vrije Universiteit Amsterdam, 
de Boelelaan 1081a, 1081 HV Amsterdam, The Netherlands; e-mail: 
{\tt dznamen@cs.vu.nl}.

\end{document}